\documentclass[sigconf]{acmart}

\AtBeginDocument{%
  \providecommand\BibTeX{{%
    \normalfont B\kern-0.5em{\scshape i\kern-0.25em b}\kern-0.8em\TeX}}}

\setcopyright{acmcopyright}
\copyrightyear{2024}
\acmYear{2024}
\acmDOI{XXXXXXX.XXXXXXX}

\acmConference[XXX'24]{the ACM XXX Conference 2024}{2024}{Singapore}
\acmPrice{15.00}
\acmISBN{978-1-4503-XXXX-X/18/06}

\usepackage{graphicx}
\usepackage{stfloats}
\usepackage{hyperref}
\usepackage{subfigure}
\usepackage{scalerel,xparse}

\graphicspath{ {./images/} }
\begin{document}

\NewDocumentCommand\emojiRAHa{}{
    \scalerel*{
        \includegraphics{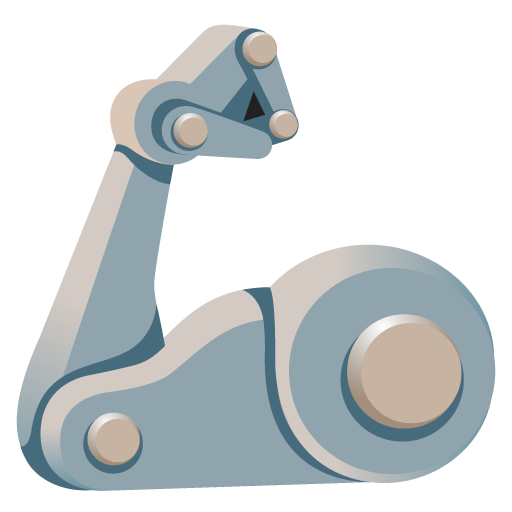}
    }{X}
}

\NewDocumentCommand\emojiRAHb{}{
    \includegraphics[scale=0.025]{images/emoji-mechanical-arm.png}
}


\title{RAH!\emojiRAHa RecSys-Assistant-Human: A Human-Centered Recommendation Framework with LLM Agents}


\author{Yubo Shu}
\affiliation{%
  \institution{School of Computer Science, Fudan University}
  \city{Shanghai}
  \country{China}}
\email{ybshu20@fudan.edu.cn}

\author{Haonan Zhang}
\affiliation{%
  \institution{School of Computer Science, Fudan University}
  \city{Shanghai}
  \country{China}}
\email{hnzhang23@m.fudan.edu.cn}

\author{Hansu Gu}
\affiliation{%
  \institution{Seattle}
  \country{United States}}
\email{hansug@acm.org}

\author{Peng Zhang}
\authornote{Corresponding author.}
\affiliation{%
  \institution{Shanghai Key Laboratory of Data Science, Fudan
University}
  \city{Shanghai}
  \country{China}}
\email{zhangpeng_@fudan.edu.cn}

\author{Tun Lu}
\authornotemark[1]
\affiliation{%
  \institution{School of Computer Science, Fudan University}
  \city{Shanghai}
  \country{China}}
\email{lutun@fudan.edu.cn}

\author{Dongsheng Li}
\affiliation{%
  \institution{Microsoft Research Asia}
  \city{Shanghai}
  \country{China}}
\email{dongshengli@fudan.edu.cn}

\author{Ning Gu}
\affiliation{%
  \institution{School of Computer Science, Fudan University}
  \city{Shanghai}
  \country{China}}
\email{ninggu@fudan.edu.cn}

\renewcommand{\shortauthors}{Yubo Shu, et al.}
\settopmatter{printacmref=false}
\begin{abstract}

The rapid evolution of the web has led to an exponential growth in content. Recommender systems play a crucial role in Human-Computer Interaction (HCI) by tailoring content based on individual preferences. Despite their importance, challenges persist in balancing recommendation accuracy with user satisfaction, addressing biases while preserving user privacy, and solving cold-start problems in cross-domain situations. This research argues that addressing these issues is not solely the recommender systems' responsibility, and a human-centered approach is vital. We introduce the \textbf{RAH} (\textbf{R}ecommender system, \textbf{A}ssistant, and \textbf{H}uman) framework, an innovative solution with LLM-based agents such as Perceive, Learn, Act, Critic, and Reflect, emphasizing the alignment with user personalities. The framework utilizes the Learn-Act-Critic loop and a reflection mechanism for improving user alignment. Using the real-world data, our experiments demonstrate the RAH framework's efficacy in various recommendation domains, from reducing human burden to mitigating biases and enhancing user control. Notably, our contributions provide a human-centered recommendation framework that partners effectively with various recommendation models. 

\end{abstract}
\maketitle

\section{Introduction}
Recommender systems hold a pivotal role in Human-Computer Interaction (HCI) by personalizing content and services to individual preferences, thereby enriching user experience and aiding in decision-making~\cite{swearingen2001beyond}. They efficiently filter information, effectively managing overload and assisting users in locating relevant content. However, there remain notable challenges. Striking the delicate balance between recommendation accuracy and user satisfaction is a fundamental objective~\cite{mcnee2006being, konstan2021human}. Addressing biases in recommendations~\cite{chen2023bias} and empowering users with control while preserving their privacy remains a pressing concern~\cite{ge2022survey}. Additionally, simplifying transitions into new domains and alleviating user burden stand as ongoing challenges~\cite{zang2022survey}, typically revealing themselves as a cold start problem. 

While much of the pioneering research primarily focuses on addressing challenges from the perspective of the recommender system, we argue that solving these issues is not the sole responsibility of recommender systems. Addressing challenges from the human perspective presents a new and promising angle. For instance, employing advanced user modeling techniques to capture user behavior and preferences allows for a delicate balance between user satisfaction and recommendation precision. Engaging users in a cooperative manner within the recommendation process enables them to define profiles, tailor preferences, and provide explicit feedback. This not only helps mitigate biases but also empowers users, enhancing their control over recommendations and protecting privacy. When confronted with the cold-start challenge, understanding user preferences and effectively generalizing them in uncharted domains can significantly alleviate the burden on users entering unfamiliar territories. These human-centered strategies represent orthogonal efforts to complement existing recommender systems.

We propose a comprehensive framework \textbf{RAH}, which stands for \textbf{R}ecommender system, \textbf{A}ssistant, and \textbf{H}uman. Within this framework, the assistant acts as an intelligent and personalized helper, leveraging LLM to learn and comprehend a user's personality from their behaviors. The assistant then provides tailored actions in line with the user's personality. Operating within this framework, RAH opens up avenues to alleviate user burden, mitigate biases, and enhance user control over recommended outcomes and personal privacy. Each assistant comprises several LLM-based agents. (1) Perceive Agent: Understands and interprets information within recommendations, including item features and user feedback implications. (2) Learn Agent: Assimilates user personalities from their behaviors and stores them in personality libraries. (3) Act Agent: Executes actions based on the learned personality, such as filtering out disliked items for the user. (4) Critic Agent: Validates if the executed action aligns with the user's preferences and analyzes adjustments to reduce discrepancies. (5) Reflect Agent: Scrutinizes and optimizes the accumulated learned personality, addressing issues like duplication and conflicts. Furthermore, we enhance our proposed assistant with the Learn-Act-Critic loop and a reflection mechanism to enhance alignment with the user. Within the Learn-Act-Critic loop, the Learn, Act, and Critic Agents work collaboratively to process user actions, refining their understanding of the user's personality. This iterative loop continues until the Act Agent accurately mirrors the learned personality, ensuring alignment with user interactions validated by the Critic Agent. Meanwhile, the reflection mechanism employs the Reflect Agent to periodically revise the learned personality, maintaining an up-to-date and accurate representation.

In our experiment, we evaluate the RAH framework using real-world data in three recommendation domains. Firstly, we observe that the Learn-Act-Critic loop and reflection mechanism significantly enhance the alignment of the assistant with the user's personality. Post-learning from users, the assistant is capable of generating proxy actions across various recommender systems, effectively reducing human burden. The second experiment demonstrates that these proxy actions lead to a notable improvement in recommender systems, achieving enhanced efficiency with reduced user interactions. Moreover, in the third part of the experiment, we investigate the use of well-learned assistants to express users' feedback on less popular items, mitigating bias within the system. Finally, we delve into additional strategies within the RAH framework to tackle human-centered concerns regarding user control. The assistant comprehends users' intentions, delivers more detailed recommended results to fulfill them, and implements control strategies to safeguard users' privacy.

\begin{figure*}[h!]
    \includegraphics[trim=0.1cm 0.4cm 0.1cm 0.2cm, clip=true, width=0.8\textwidth]{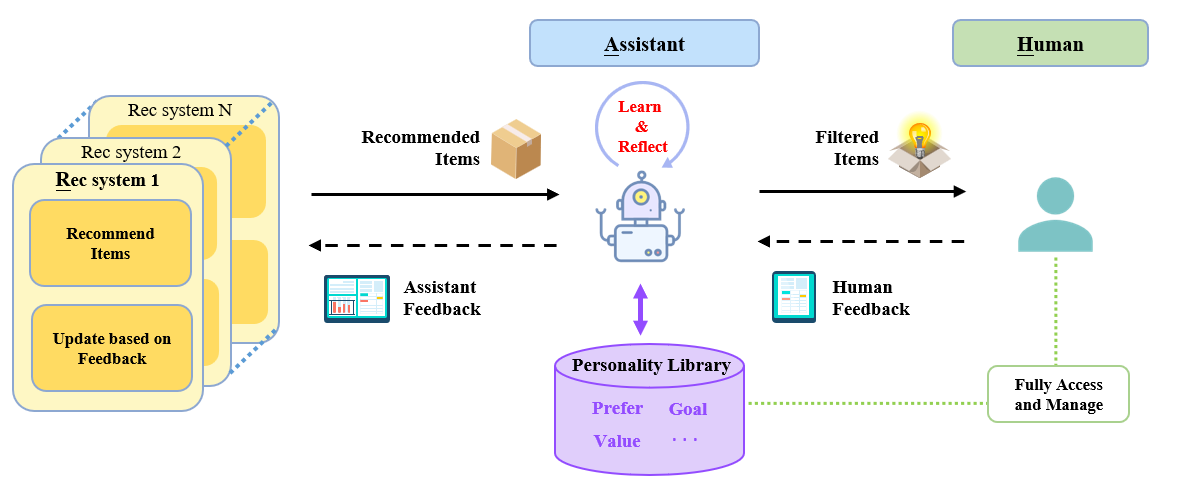}
    \caption{The figure demonstrates an overall view of the RAH framework. Core workflows can be divided into RecSys$\rightarrow$Assistant$\rightarrow$Human(the black solid arrow) and Human$\rightarrow$Assistant$\rightarrow$RecSys(the black dotted arrow).}
    \label{figure:framework}
    \centering
\end{figure*}

Our contributions can be summarized as follows:
\vspace{-0.1cm}
\begin{itemize}
  \item We utilize LLM from the human perspective and propose a more human-centered recommendation framework, RAH.
  \item Within the RAH framework, our assistant is designed with the Learn-Act-Critic loop and a reflection mechanism to achieve a nuanced understanding and alignment with user personalities.
  \item Through experimentation, we validate the RAH framework's performance in addressing recommendation challenges partnered with various recommendation models, including cold-start in cross-domain recommendation, popularity bias, and user control and privacy.
\end{itemize}

\section{RAH (RecSys-Assistant-Human)}

\subsection{Overall}

The principle behind RAH's design is taking a human-centered approach to address recommender system challenges. As shown in Figure~\ref{figure:framework}, RAH comprises three components - the recommender system, the intelligent assistant, and the human user. Unlike traditional recommendations solely between systems and users, RAH introduces an assistant as an intermediary. This assistant acts as a personalized helper for the user. It utilizes large language models (LLMs) to comprehend user personalities based on their behaviors. The assistant then provides actions tailored to each user's personality.

Within this framework, the assistant facilitates two key workflows:

\textbf{RecSys$\rightarrow$Assistant$\rightarrow$Human}
This workflow focuses on the assistant filtering personalized recommendations for the end user, as shown by the solid black arrow in Figure~\ref{figure:framework}.

\begin{itemize}
    \item Recommender systems initially generate candidate items spanning different domains such as books, movies, and games. 
    \item The assistant aggregates these cross-domain recommendations. It retrieves the user's learned personality from its memory. Using the user's personality profile, the assistant further filters the candidate items to create a tailored list.
    \item Finally, the user receives a unified personalized set of filtered recommendations from the assistant.
\end{itemize}

To enable effective filtering across diverse items, the assistant incorporates powerful LLMs. They provide the reasoning skills and real-world knowledge needed to comprehend various item features.

\textbf{Human$\rightarrow$Assistant$\rightarrow$RecSys}
This workflow enables the assistant to learn from user feedback and accordingly tune recommender systems, as depicted by the dotted black arrow in Figure~\ref{figure:framework}.

\begin{itemize}
    \item The user first provides feedback on items, e.g., indicating ``Like" or ``Dislike", and the assistant receives this initial feedback instead of the recommender systems.
    \item The assistant will then start to learn the user's personality from the user's feedback. 
    \item Lastly, the assistant will process the user's feedback into the assistant's feedback. This allows it to selectively forward user preferences to recommender systems.
\end{itemize}

By introducing an intermediary assistant focused on the human, RAH opens up new possibilities to address human-centered challenges. The assistant's capabilities in learning and acting upon user personalities strengthen these human-centered aspects. It facilitates key functionalities like mitigating user burden and bias while enhancing user control and privacy.

\subsection{Human-Centered Design Goals}

As stated earlier, the key goal of RAH is to address human-centered challenges in recommender systems. This subsection introduces three pivotal design goals for addressing human-centered challenges. (Our methods to achieve the design goals can be found in Section~\ref{sec:human_centered_approaches})

\textbf{Reduce User Burden.} 
In recommendation, the user burden can come from the initial interactions in a new domain and the redundant feedback across domains. In the RAH framework, the assistant should serve as a personal helper to reduce user burden in multiple ways. In both a single domain and across domains, the assistant should comprehend user tendencies from limited interactions and learn a unified user personality. The assistant should be able to express a unified personality to new recommender systems, alleviating the cold start issue and reducing user burden. Besides, the assistant should provide proxy feedback to refine recommender systems, minimizing unnecessary user interactions.

\textbf{Mitigate bias.} 
Biased recommended results can cause unfairness problems and harm the user experience. In the RAH framework, we design the assistant to represent users, generating more feedback on unseen items and thus mitigating the user's selection bias.

\textbf{Enhance User Control.} 
Considering the pattern that the recommender system actively interacts with users, it is necessary to address user control in recommendation \cite{sulikowski2018human, shin2020users}. However, the majority of the current recommender systems are uncontrollable, and users can only passively receive the recommendation results \cite{ge2022survey}. Therefore, in the RAH framework, the assistant should enhance user control of the recommendation results they receive and what the recommender systems learn about them, such as non-privacy data.

\begin{figure*}[h]
    \centering
    \includegraphics[trim=0.1cm 0.1cm 0.1cm 0.1cm, clip=true, width=0.9\textwidth]{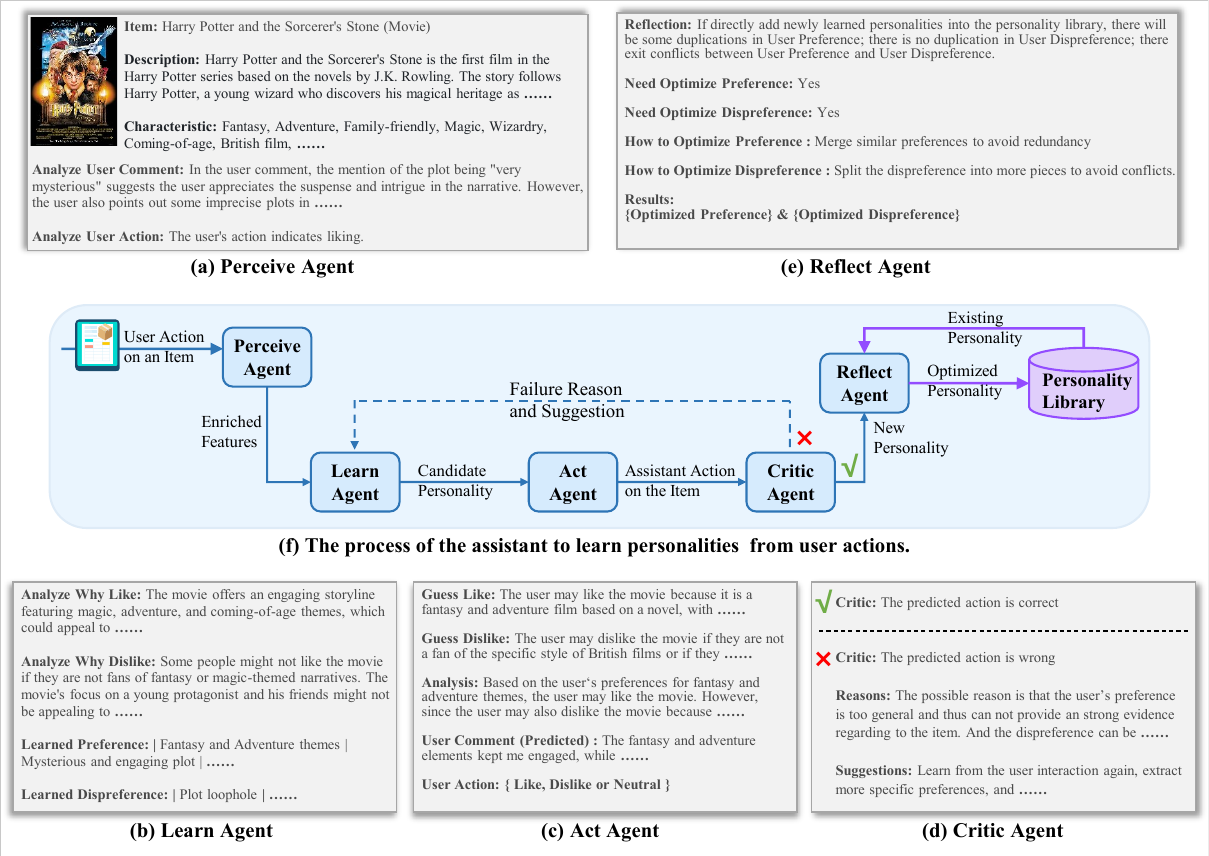}
    \caption{The components of the assistant and their work pattern.}
    \label{figure:assistant}
    \centering
\end{figure*}

\section{Assistant}

In this section, we first provide an overview of the assistant's components and inner mechanisms. We then elaborate on how the assistant achieves human-centered goals.

\subsection{Components}

\subsubsection{\textbf{Perceive Agent}}

The Perceive Agent functions as the initial processing point for incoming information. Specifically, in the context of recommendations, its primary task is to augment the features associated with a given item, thereby enhancing the assistant's overall comprehension. For instance, when provided with a movie name, the Perceive agent can supply additional relevant information about the movie. As illustrated in Figure~\ref{figure:assistant}(a), this additional information generally consists of two components: (1) a concise description of the item, such as a plot summary of the movie, and (2) a set of specific attributes related to the item, like the movie tags. Additionally, this information enriched by the Perceive agent can further aid other agents, such as assisting the Learn Agent in extracting personalities from user behaviors.

\subsubsection{\textbf{Learn Agent}}

The Learn Agent's mission is to identify human personalities based on interactions with items, such as ``Like", ``Dislike", and user ratings. Drawing inspiration from established research in recommender systems \cite{lee2019melu, geng2022recommendation, sanner2023large}, we conceptualize human personalities as a combination of likes and dislikes. In our implementation, we input items, human feedback on items, and insights from the Perceive Agent into the Learn Agent. As depicted in Figure~\ref{figure:assistant}(b), the Learn Agent then generates the learned preferences in response to positive feedback and the dislikes for negative feedback. Moreover, instead of direct learning, we require the agent to address two key questions: ``Why might some individuals like the item?" and ``Why might some individuals dislike the item?" These responses aid the agent in filtering out invalid characteristics and promoting a more nuanced understanding of personalities.

\subsubsection{\textbf{Act Agent}}

The Act Agent is responsible for generating actions based on the learned personality. The Act Agent receives an item's information and a user's personality as input. Subsequently, it generates a predicted action, such as "Like" when the item aligns with the user's preferences and "Dislike" when it aligns with their dislikes. As shown in Figure~\ref{figure:assistant}(c), we incorporate a chain-of-thoughts \cite{wei2022chain} approach in our implementation: (1) hypothesizing reasons for potential preference or dislikes towards the item, (2) analyzing the likely perception of the item by a human with the given personality, (3) simulating comments on the item from the perspective of the human \cite{lin2018explainable, zhang2020explainable}, and finally, (4) predicting the human's reaction to the item, categorized as either ``like" or ``dislike."

\subsubsection{\textbf{Critic Agent}}

The core function of the Critic Agent is to evaluate the correctness of actions predicted by Act Agents. A match between the predicted action and the ground truth action (true user actions) suggests that the learned personality model aligns with the user. However, in cases of incorrect predictions, the Critic Agent not only identifies the discrepancy between predictions and labels but also analyzes potential reasons for the failure to facilitate corrective measures. As depicted in Figure~\ref{figure:assistant}(d), this process can be compared to a code compiler detecting a bug in code and generating an error log, enabling the programmer to identify and rectify the issue. As a result, the reasons for failure are conveyed to the Learn Agent, prompting a reevaluation of previous attempts and a relearning of the personality~\cite{wang2023voyager}. This iterative collaboration between the Learn, Act, and Critic Agents enhances the inference of human personality based on observed actions.

\subsubsection{\textbf{Reflect Agent}}

The Reflect Agent's role is to periodically review the learned personality. As illustrated in Figure~\ref{figure:assistant}(e), the Reflect Agent's input comprises the combination of newly acquired learned personality and existing personalities. The Reflect Agent then evaluates the combined personalities, identifying duplicate likes, duplicate dislikes, and conflicts between likes and dislikes. The rationale behind employing the Reflect Agent is to ensure the rationality of the learned personalities throughout the continuous learning process.

\subsection{Enhance Alignment}

Given the critical importance of aligning with the user, we further implement a Learn-Act-Critic loop and a reflection mechanism to reinforce this alignment.

\textbf{Learn-Act-Critic Loop.} As shown in Figure~\ref{figure:assistant}(f), our Learn Agent collaborates with the Act and Critic Agents in an iterative process to grasp the user's personality. Upon receiving user action or feedback, the Learn Agent extracts an initial personality as a candidate. Then, the Act Agent utilizes this candidate as input to predict the user's actual action in reverse. The Critic Agent then assesses the accuracy of this prediction. If the prediction proves inaccurate, the Critic Agent delves into the underlying reasons and offers suggestions for corrections. The Learn Agent then incorporates these suggestions, refining the candidate's personality until it meets the Critic Agent's evaluation.

\textbf{Reflecting on personality.} To attain more accurate and comprehensive personalities, the assistant must seamlessly integrate the newly acquired personality with existing ones, rather than merely accumulating them. Inspired from \cite{park2023generative}, our reflection mechanism addresses issues arising from duplication and conflicts in learned personalities (preferences and aversions). Regarding duplication, the assistant can effortlessly merge duplicates without requiring additional information. However, handling conflicts may require a more delicate strategy. The Reflect Agent initiates by deconstructing conflicting traits into finer details to minimize overlaps. If conflicts persist after this step, the Reflect Agent formulates queries for users, seeking their input to resolve the conflicts.

\subsection{Human-Centered Approaches}
\label{sec:human_centered_approaches}
In this section, we discuss key human-centered approaches employed within the RAH framework to reduce user burden, mitigate biases, and enhance user control.

\textbf{Reduce user burden.} 
The assistant reduces user burden through its learning and acting capabilities. It employs the Learn Agent to learn a unified user personality from diverse domain interactions in the user's history. This unified personality is then extrapolated across domains using the Act Agent, resulting in personalized proxy feedback to instruct recommender systems. This process helps users avoid abundant interactions and thus reduces user burden. Within a single domain, the assistant utilizes powerful LLMs to comprehend user personalities with fewer actions. Across domains, this unified personality alleviates the 'cold start' issue and reduces the initial feedback burden. Additionally, the assistant can analyze user behavior across mixed domains, gradually constructing a more comprehensive personality that aligns better with the user.

\textbf{Mitigate bias.} 
To mitigate bias, the assistant leverages the Act Agent to act on items and generate proxy feedback. Human feedback, limited by time and energy, tends to be biased towards popular or seen items. The Act Agent addresses this limitation by offering expanded feedback on less popular or unseen items, thus reducing selection bias. This broader interaction history leads to less biased recommendations from the recommender systems. The Action Agent, based on LLMs, provides nuanced feedback, such as proxy comments, allowing for a deeper understanding of explicit user preferences. This enables recommender systems to focus on genuine user preferences rather than simply fitting to the training data, thus reducing inference bias.

\textbf{Enhance user control.}
Different from the traditional framework consisting of users and a remote recommendation system, the assistant is designed to prioritize users' intentions and objectives. With the integration of LLMs, the assistant can operate on personal devices~\cite{mlc-llm}, empowering users and providing a more human-centered experience. The Act Agent plays a crucial role in enhancing user control through content filtering and tailored recommendations:

\begin{itemize}
    \item \textbf{Control recommendation results:} 
    Equipped with LLM, the Learn Agent comprehends complex human intentions effectively. The Act Agent then filters items and tailors recommender systems to ensure recommended results align with user intentions. For instance, if a user instructs the assistant to exclude horrifying elements, the assistant filters out such movies, books, and games from recommendations and generates proxy actions such as ``Dislike" for items containing these elements.

    \item \textbf{Control privacy:} 
    Beyond operating on personal devices, the assistant employs strategies to enhance privacy and personalized recommendations. The assistant limits data sharing with recommender platforms and employs obfuscation strategies, such as providing obfuscated proxy feedback to mask a user's identity. For example, if a patient expresses interest in a treatment-related book, the assistant could provide extra proxy feedback, such as ``Likes Professional Medical Literature", to the recommender system, thereby masking the patient's identity and suggesting they might be a medical professional. In response, the recommender system might suggest a mix of treatment-focused books and advanced medical literature. The assistant then uses the Act Agent to filter out the specialist literature, presenting only the relevant treatment-related options to the user. This strategy ensures privacy while delivering personalized recommendations tailored to the user's needs.
    
\end{itemize}

\section{Experiments Setting}

In this section, we outline the specifics of our experiments and dataset preparation. Our evaluation of the RAH framework involves three experiments to assess: (1) the assistant's alignment with the user preference. (2) the performance of reducing user burden among various domains, and (3) the assistant's capability to mitigate bias. For all experiments, we utilize the GPT-4-0613 version of the LLM from OpenAI in our assistant.

Our datasets are sourced from three domains on Amazon: Movies, Books, and Video Games. Following the guidelines of previous research \cite{liu2022collaborative}, we initially filter out users and items with fewer than five interactions. We then retain users who have interactions in more than one domain, allowing us to additionally evaluate RAH's performance in cross-domain situations (e.g., Movie\&Book). 
Subsequently, to strike a balance between GPT-4 API calls and the training demands of the recommender system, we split the dataset into two parts:

\begin{itemize}
    \item \textbf{Cross1k.} We randomly select 1,000 users from the processed data, capturing their interactions to form a concise dataset. For these users, 1,000 personalized LLM-based assistants are created to learn from and act to them individually. For the following experiments, we further partition the interactions of Cross1k into three sets (Learn Set, Proxy Set, and Unseen Set) using an equal ratio of 1:1:1.
    \item \textbf{Cross221k.} The rest of the dataset includes 221,861 users and 4,624,903 interactions, and it can be used for training a stable recommender system without the challenges tied to insufficient training data. 
\end{itemize}
The statistics of Cross1k and Cross221k can be found in Appendix~\ref{app:datasets}.
To test RAH's role in reducing bias, we follow the protocols with previous de-bias research \cite{bonner2018causal, zheng2021disentangling, wan2022cross} to simulate unbiased data for offline evaluation by sampling interactions according to the propensity scores of items.

\begin{figure*}[h]
    \centering
    \includegraphics[trim=0.05cm 0.05cm 0.05cm 0.02cm, clip=true, width=0.8\textwidth]{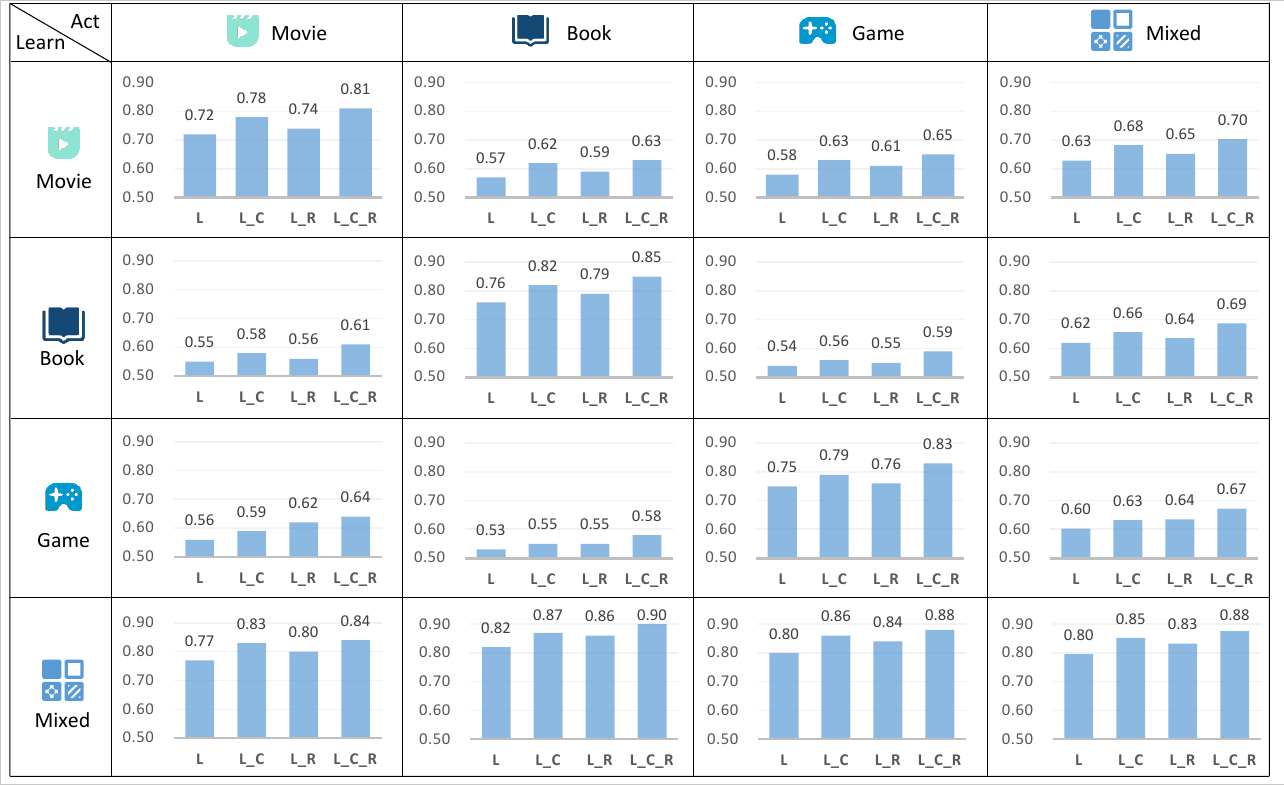}
    \caption{Performance evaluation of the assistant's ability to align with users across singular, cross, and mixed domains. Histogram values represent the F1-Score against user actions. \textbf{L} for Learn Only, \textbf{C} for using Learn-Act-Critic loop, and \textbf{R} for the reflection mechanism.} 
    \label{figure:e1_learn}
    \centering
\end{figure*}

\section{Results and Discussion}
In this section, we first showcase our experimental results focusing on alignment, burden reduction, and bias mitigation. Subsequently, we explore case studies emphasizing enhanced user control over recommended outcomes and personal privacy.

\subsection{Assistants' Alignment with Users}

For the first alignment-focused experiment, we task the assistant with assimilating personalities from the Learn Set and then generating proxy actions for items within the Proxy Set in Cross1k.
In order to evaluate our assistant's alignment with users, an intuitive measure is whether an assistant can take consistent actions with a user. Therefore, the evaluation process is: (1) We instruct the assistant to extract users' personalities from their interactions in the Learn Set, such as ratings and comments on items. (2) The assistant is then tasked with predicting actions on items in the Proxy Set. We then examine if these predicted actions align with the actual behaviors of users.

To gain a more comprehensive evaluation, we conduct the experiment to include both cross-domains and mixed domains. For comparison, we have four tasks for personality learning:
\begin{itemize}
  \item Learn Only: We directly append learned new likes or dislikes into users' personalities without Critic Agent or Reflect Agent. 

  \item Learn+Reflect: After appending new likes or dislikes to users' personalities, we employ the reflection mechanism to resolve potential duplication and conflicts. 

  \item Learn+Critic: After acquiring new likes or dislikes from a particular user action, we input the new likes or dislikes and assess if the Act Agent can accurately infer the original user action in reverse. If not successful, the assistant should attempt another Learn-Act-Critic loop. 

  \item Learn+Critic+Reflect: Both the Learn-Act-Critic loop and reflection mechanism are engaged for optimization.
\end{itemize}

Figure~\ref{figure:e1_learn} presents the F1-score of the personality learning experiment. Overall, compared with Learn Only, either the learn-act-critic loop or reflection mechanism is helpful in aligning with users. Moreover, their combined application yields even more significant improvements.

Learning and acting within the same domain yields better results compared to cross-domain operations. Furthermore, the results demonstrate that learning from a mixed domain outperforms learning from any single domain, such as movies, books, or games when considered independently. This suggests that LLM-based assistants possess the capability to reason and extrapolate users' personalities across different domains.

\subsection{Reduce Human Burden}
\label{sec:reduce_human_burden}
\begin{table*}[h]
\centering
\caption{The performance of proxying user feedback and adjusting recommender systems.}
\label{table:rec_experiment}
\footnotesize
\begin{tabular}{llllllllll}
\toprule
\textbf{Method}  &\textbf{Assistant}& \multicolumn{2}{c}{\textbf{Movie}}& \multicolumn{2}{c}{\textbf{Book}}& \multicolumn{2}{c}{\textbf{Game}}& \multicolumn{2}{c}{\textbf{Mixed}}\\
\toprule
 & & NDCG@10& Recall@10& NDCG@10& Recall@10& NDCG@10& Recall@10& NDCG@10&Recall@10\\
\midrule
LightGCN &No&0.5202 & 0.5142 & 0.1283 & 0.1439 & 0.3459 & 0.4309 & 0.3403 &0.1696 \\
LightGCN &Yes& 0.5524(+0.0322)& 0.5339(+0.0197)& 0.1830(+0.0547)& 0.1912(+0.0473)& 0.4330(+0.0871)& 0.4974(+0.0665)& 0.4058(+0.0655)&0.2033(+0.0337)\\
\midrule
PLMRec&No
&0.0993 & 0.1316 & 0.0092 & 0.0143 & 0.3693 & 0.4630 & 0.1075 &0.0656 \\
 PLMRec&Yes& 0.1200(+0.0207)& 0.1692(+0.0376)& 0.0162(+0.0070)& 0.0197(+0.0054)& 0.3981(+0.0288)& 0.4790(+0.0160)& 
0.1378(+0.0303)&0.0766(+0.0110)\\
\midrule
 FM&No
& 0.3492 & 0.3871 & 0.1216 & 0.1299 & 0.2917 & 0.3586 & 0.2421 &0.1262 \\
 FM&Yes& 0.3919(+0.0427)& 0.4257(+0.0386)& 0.1474(+0.0258)& 0.1603(+0.0304)& 0.2937(+0.0020)& 0.3624(+0.0038)& 
0.2549(+0.0128)&0.1340(+0.0078)\\
\midrule
 MF&No
& 0.3737 & 0.4450 & 0.1143 & 0.1275 & 0.2074 & 0.2622 & 0.1933 &0.1054 \\
 MF&Yes& 0.4300(+0.0563)& 0.4781(+0.0331)& 0.1520(+0.0377)& 0.1593(+0.0318)& 0.2998(+0.0924)& 0.3706(+0.1084)& 0.2651(+0.0718)&0.1487(+0.0433)\\
\midrule
 ENMF&No
& 0.4320 & 0.3953 & 0.0994 & 0.0997 & 0.0652 & 0.1036 & 0.2630 &0.1227 \\
 ENMF&Yes& 0.5200(+0.0880)& 0.4831(+0.0878)& 0.1224(+0.0230)& 0.1217(+0.0220)& 0.0788(+0.0136)& 0.1247(+0.0211)& 0.3224(+0.0594)&0.1531(+0.0304)\\
\midrule
 NeuralMF&No
& 0.4720 & 0.4878 & 0.1364 & 0.1385 & 0.2160 & 0.2704 & 0.2891 &0.1507 \\
 NeuralMF&Yes& 0.4856(+0.0136)& 0.4906(+0.0028)& 0.1631(+0.0267)& 0.1658(+0.0273)& 0.3507(+0.1347)& 0.4086(+0.1382)& 0.3451(+0.0560)&0.1742(+0.0235)\\
\midrule
 ItemKNN&No
& 0.1211 & 0.1035 & 0.0889 & 0.0694 & 0.2242 & 0.3074 & 0.1657 &0.0790 \\
 ItemKNN&Yes& 0.2131(+0.0920)& 0.1860(+0.0825)& 0.1517(+0.0628)& 0.1171(+0.0477)& 0.2660(+0.0418)& 0.3125(+0.0051)& 0.2567(+0.0910)&0.1170(+0.0380)\\
\bottomrule
\end{tabular}
\end{table*}

In the second experiment, we connect the assistant with traditional recommender systems within the RAH framework. To evaluate whether the assistant can reduce user burden, we measure how effectively the assistant can represent users and provide proxy feedback to calibrate the recommender systems using the RAH framework. We perform comparison experiments for various recommendation algorithms, both with and without assistants.

\textbf{Without assistants}, we train recommendation algorithms on Cross221k and the Learn Set of Cross1k. Lastly, we calculate the recommendation metric on the Unseen Set. \textbf{With assistants}, we initially use assistants to learn each user's personality on Learn Set and let the assistant make proxy feedback on Proxy Set (same as the first experiment). Then we train recommendation models on Cross221k, Learn Set and the assistant's proxy feedback, and likewise test on Unseen Set. The involved recommendation algorithms are as follows:
\begin{itemize}
  \item LightGCN\cite{he2020lightgcn}: A model that enhances recommender systems by simplifying neighborhood aggregation, and learns embeddings through linear propagation on the interaction graph. 
  \item PLMRec\cite{wu2021empowering}: A recommendation model that uses PLMs like Bert to embed the content of items for deeper semantic mining. 
  \item FM\cite{rendle2010factorization}: Model that combines SVM advantages with factorization models, using factorized parameters to model interactions in sparse data. 
  \item MF\cite{koren2009matrix}: Use matrix factorization techniques for recommendation systems to generate product recommendations by using historical data.
  \item ENMF\cite{chen2020efficient}: Based on simple neural matrix factorization, it optimizes model parameters from the entire training data without sampling. 
  \item NeuralMF\cite{he2017neural}: A framework that uses deep neural networks modeling collaborative filtering based on implicit feedback and user-item feature interactions. 
  \item ItemKNN\cite{deshpande2004item}: An item-based Top-N recommendation algorithm that uses item similarities to determine the recommendation set. 
\end{itemize}

Table~\ref{table:rec_experiment} presents the results of our comparison. The data suggest that, conditioned on an equal number of user interactions, the performance of various recommender systems can be improved when the assistant is integrated. Namely, after learning user personalities, the assistant can effectively calibrate recommender systems using proxy feedback. These outcomes resonate with the non-invasion design of the RAH framework. The assistant preserves the inherent pattern between the recommender system (which recommends items and gathers feedback) and the user (who receives recommendations and provides feedback). As a result, the RAH framework demonstrates remarkable adaptability across various recommender systems.

\subsection{Mitigate Bias}

\begin{table}[h]
\centering
\caption{The performance of alleviating bias. }
\label{table:e3_bias}
\small
\begin{tabular}{lcc}
\toprule
\textbf{Method} & \textbf{NDCG@10}& \textbf{Recall@10}
\\
\midrule
MF                  & 0.1835& 0.2085\\
MF+IPS              & 0.2148& 0.2424\\
MF+RAH              & 0.5017& 0.4326\\
MF+IPS+RAH          & \textbf{0.5196}& \textbf{0.4554}\\
\bottomrule
\end{tabular}
\end{table}

\begin{figure}
\centering
\subfigure[Control Recommendation Results]{\label{fig:case_result}
\includegraphics[width=0.48\textwidth]{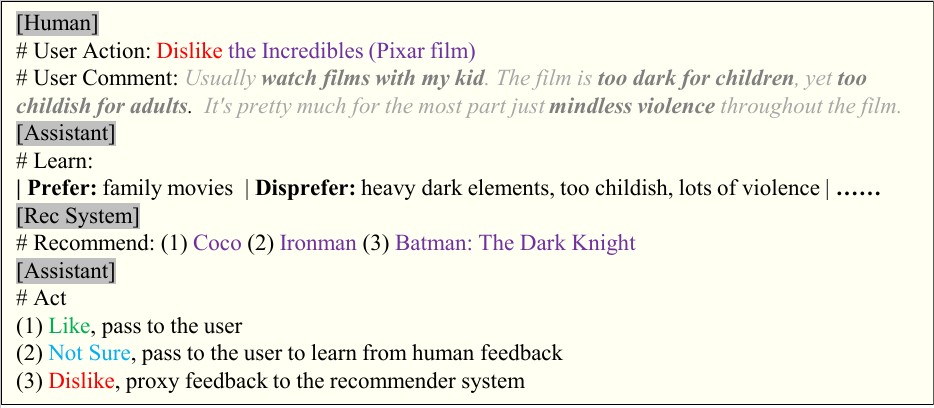}}
\subfigure[Control Personal Privacy]{\label{fig:case_privacy}
\includegraphics[width=0.48\textwidth]{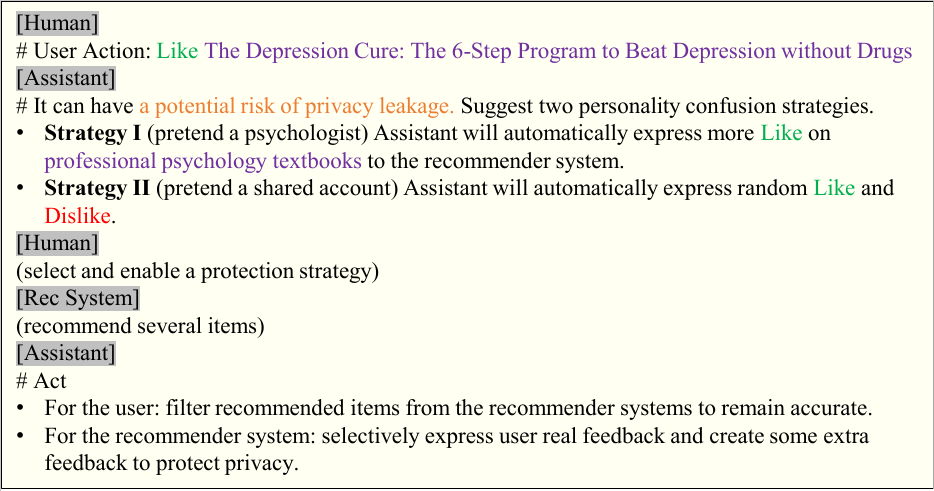}}
\caption{The case study.}
\end{figure}

In the RAH framework, the assistant provides an opportunity to address the bias problem. The above experiments demonstrate the capability of assistants to learn from user actions and make proxy feedback on items. Therefore, the assistant can also represent human users to provide proxy feedback on unpopular items and alleviate the bias in the system. To conduct the experiment, we select unpopular items (associated with less than ten reviews) in the Cross1k dataset and randomly sample user assistants to make proxy feedback on unpopular items until these items own no less than ten reviews. For comparison, we also compare a de-biasing method, Inverse Propensity Scoring (IPS) \cite{schnabel2016recommendations}. The IPS method in recommender systems adjusts for selection bias by reweighting observed data based on the likelihood of an item being recommended. 

Subsequently, we evaluate the performance on simulated unbiased test data derived from sampling. Specifically, the probability of sampling a user-item interaction is formulated to be inversely proportional to the frequency of the involved item \cite{wan2022cross}. Table~\ref{table:e3_bias} shows that both IPS and RAH are effective in mitigating bias compared with the baseline. Remarkably, when combined, the IPS and RAH approach emerges as a particularly robust de-biasing technique \cite{chen2023bias}, showing a greater efficacy in bias reduction.

\subsection{Increase User Control}

\subsubsection{Control Recommendation Results}
The first case, as illustrated in Figure \ref{fig:case_result}, demonstrates how the assistant can enhance user control over recommended results. In this case, since the user often watches movies with a child, the user expresses dissatisfaction with the movie \textit{The Incredibles} citing reasons such as it being "too childish for adults" and "too dark for children." From this feedback, the assistant discerns that the user favors family movies that strike a balance in content, avoiding extremes in themes. 

Subsequently, the recommender system suggests three movies: \textit{Coco}, \textit{Ironman}, and \textit{Batman: The Dark Knight}. Leveraging the reasoning capabilities and real-world knowledge of LLMs, the assistant can make informed decisions on items to align with user intentions. For \textit{Coco}, the assistant identifies it as a likely match for the user due to its family-friendly nature and passes the recommendation to the user. Regarding \textit{Ironman}, the assistant, uncertain of its suitability, also passes this recommendation to the user, seeking additional feedback. In contrast, \textit{Batman: The Dark Knight}, known for its dark and potentially violent content, is deemed possibly unsuitable based on the user's preferences. The assistant decides to ``Dislike" this recommendation on behalf of the user, supplying proxy feedback to the recommender system for future refinement.

\subsubsection{Control Privacy}
The second case, depicted in Figure \ref{fig:case_privacy}, highlights how the assistant can bolster user control concerning personal privacy. In this case, A user expresses interest in a specific book titled \textit{The Depression Cure: The 6-Step Program to Beat Depression without Drugs}. The assistant identifies that such an action might lead to potential privacy leakages--expressing a preference for content on mental health might disclose sensitive information about the user. The assistant offers two personality confusion strategies to help control privacy.

\textbf{Strategy I (Pretend a Psychologist)}: The assistant, mimicking the behavior of a psychologist, will express more "Like" on professional psychology textbooks within the recommender system. This action serves to dilute the user's preference, making it ambiguous whether the original interest in the depression-related book was due to personal reasons or professional curiosity.

\textbf{Strategy II (Pretend a Shared Account): } The assistant will automatically generate a mix of random likes and dislikes. This strategy gives the impression of multiple users sharing on a single account, thereby obfuscating individual preferences and adding a layer of ambiguity to the user's actions.

If the user adopts one strategy, the assistant selectively provides real user feedback and creates additional feedback, further protecting privacy. Besides, the assistant can also filter items from the recommender system to ensure that recommendations remain personalized despite the noise introduced by the selected strategy.

\section{Related Work}

\subsection{Human-Centered Recommendation}
The human-centered recommender system \cite{konstan2021human} focuses on understanding the characteristics and complex relationships between the recommender system and users in the recommendation scenario. Unlike the "accuracy-only" approach in traditional recommender systems, the human-centered recommender system pays more attention to user experience, taking user satisfaction and needs as optimization goals, such as privacy protection. Recent works have shown that this field has attracted researchers from social sciences and computational fields to participate in research together. \cite{yang2022practical} proposed a new federal recommendation framework called Federal Mask Matrix Factorization (FedMMF), which can protect data privacy in federal recommender systems without sacrificing efficiency and effectiveness. EANA \cite{ning2022eana} improves the training speed and effectiveness of large-scale recommender systems while protecting user privacy through an embedding-aware noise addition method. \cite{zhang2023user} proposed a new human-centered dialogue recommendation method, which provides more helpful recommendations to users by understanding and adapting to user needs during the dialogue process.

\subsection{LLM For Recommendation}
Large Language Models (LLMs) in Natural Language Processing (NLP) are now employed in recommender systems due to their vast knowledge and logical reasoning. LLMs for Recommendation (LLM4Rec) are mainly used in two ways: enhancing features and directly recommending. The first approach leverages LLMs for feature extraction, enhancing traditional systems. Notable works include encoding news\cite{zhang2021unbert,wu2021empowering,wu2022mm,yu2021tiny,liu2022boosting} and tweets\cite{zhang2022twhin} for recommendations. The second approach forms input sequences for LLMs, letting them directly recommend. \cite{liu2023chatgpt,Wang2023ZeroShotNR} relied on prompts for recommendations. \cite{bao2023tallrec} proposed a two-stage method: fine-tuning LLMs with recommendation data and then using them for recommendations. Works like \cite{gao2023chat,friedman2023leveraging,wang2022towards} delved into LLM's role in conversational recommender systems.

\subsection{LLM-based Agent}
With the emergence of Large Language Models (LLMs), their Autonomy, Reactivity, and Pro-activeness have brought hope and made some progress in the realization of intelligent agents \cite{xi2023rise}. This is a system that can engage in dialogue, complete tasks, reason, and exhibit a certain degree of autonomous action. Work \cite{park2023generative} has demonstrated the feasibility of LLM-based Agents by building an intelligent town supported by LLMs, showing that LLM-based Agents have strong credibility and adaptability. Work \cite{wang2023voyager} has built an LLM-Based Agent on the Minecraft game platform and proposed an iterative prompt mechanism of environmental feedback → execution error → self-verification, proving that LLM-based Agents have lifelong learning ability in scenarios and strong generalization ability to solve new tasks. Similarly, work \cite{sumers2023cognitive} divides the LLM-based Agent into three modules: control end, perception end, and action end from the perspective of cognitive science. Work \cite{liu2023training} proposes a training paradigm that allows LLM to learn social norms and values from simulated social interactions.

\section{Conclusion and Future Work}

From the perspective of humans, we introduce the RAH framework for recommendations, incorporating the design of the assistant using LLM Agents. Our experiments highlight the efficacy of the Learn-Act-Critic loop and reflection mechanisms in enabling the assistant to align more closely with user personalities. Besides, we evaluate the RAH framework on different recommender systems in reducing user burden and find the generalization capability of the framework, which echoes the non-invasion role of the assistant. Additionally, we measure the assistant's capability to provide proxy feedback on unpopular items to mitigate selection bias. Finally, we explore potential solutions to increase user control of recommended results and personal privacy through the assistant.

One constraint of our current approach is its reliance on offline evaluations. In the future, we plan to conduct online assessments of the RAH framework, focusing on the sustained influence of the assistant on users and recommender systems. Moreover, we will explore the collaborative relationship between the assistant and humans, such as whether personalities learned from subjective tasks like recommendations can be translated into content creation scenarios that align with user preferences.


\bibliographystyle{ACM-Reference-Format}
\bibliography{sample-base}

\section{Appendices}
\subsection{The statistics of datasets}
\label{app:datasets}
The number of users, items and interactions in different domains for both Cross1k and Cross221k.

\begin{table}[h!]
\centering
\caption{Cross1k.}
\begin{tabular}{lccc}
\hline
\textbf{Domain} & \textbf{\#Users} & \textbf{\#Items} & \textbf{\#Interactions} \\
\hline
Movie & 1,045& 10,679& 21,024\\
Book & 1,046& 20,159& 24,035\\
Game & 1,044& 8,984& 17,169\\
\hline
\end{tabular}
\end{table}

\begin{table}[h!]
\centering
\caption{Cross221k.}
\begin{tabular}{lccc}
\hline
\textbf{Domain} & \textbf{\#Users} & \textbf{\#Items} & \textbf{\#Interactions} \\
\hline
Movie & 221,861& 49,791& 2,313,890\\
Book & 94,407& 12,898 & 2,240,010\\
Game & 7,149& 12,196& 71,003\\
\hline
\end{tabular}
\end{table}

\begin{table*}[b]
\centering
\caption{The performance of proxying user feedback and adjusting recommender systems with the additional comparison.}
\label{table:expand_e2}
\footnotesize
\begin{tabular}{lllllllll}
\toprule
\textbf{Method}  & \multicolumn{2}{c}{\textbf{Movie}}& \multicolumn{2}{c}{\textbf{Book}}& \multicolumn{2}{c}{\textbf{Game}}& \multicolumn{2}{c}{\textbf{Mixed}}\\
\toprule
 & NDCG@10& Recall@10& NDCG@10& Recall@10& NDCG@10& Recall@10& NDCG@10&Recall@10\\
\midrule
LightGCN&0.5202 & 0.5142 & 0.1283 & 0.1439 & 0.3459 & 0.4309 & 0.3403 &0.1696 
\\
LightGCN-Random& 0.5341(+0.0139)& 0.5240(+0.0098)& 0.1527(+0.0244)& 0.1711(+0.0272)& 0.4163(+0.0704)& 0.4934(+0.0625)& 0.3790(+0.0387)&0.1900(+0.0204)
\\
 LightGCN-Assistant& 0.5524(+0.0322)& 0.5339(+0.0197)& 0.1830(+0.0547)& 0.1912(+0.0473)& 0.4330(+0.0871)& 0.4974(+0.0665)& 0.4058(+0.0655)&0.2033(+0.0337)\\
\midrule
PLMRec&0.0993 & 0.1316 & 0.0092 & 0.0143 & 0.3693 & 0.4630 & 0.1075 &0.0656 
\\
 PLMRec-Random& 0.1171(+0.0178)& 0.1610(+0.0294)& 0.0149(+0.0057)& 0.0181(+0.0038)& 0.3964(+0.0271)& 0.4743(+0.0113)& 
0.1346(+0.0271)&0.0739(+0.0083)
\\
 PLMRec-Assistant& 0.1200(+0.0207)& 0.1692(+0.0376)& 0.0162(+0.0070)& 0.0197(+0.0054)& 0.3981(+0.0288)& 0.4790(+0.0160)& 0.1378(+0.0303)&0.0766(+0.0110)\\
\midrule
 FM& 0.3492 & 0.3871 & 0.1216 & 0.1299 & 0.2917 & 0.3586 & 0.2421 &0.1262 
\\
 FM-Random& 0.3897(+0.0405)& 0.4200(+0.0329)& 0.1443(+0.0227)& 0.1561(+0.0262)& 0.2903(-0.0014)& 0.3529(-0.0057)& 
0.2533(+0.0112)&0.1336(+0.0074)
\\
 FM-Assistant& 0.3919(+0.0427)& 0.4257(+0.0386)& 0.1474(+0.0258)& 0.1603(+0.0304)& 0.2937(+0.0020)& 0.3624(+0.0038)& 0.2549(+0.0128)&0.1340(+0.0078)\\
\midrule
 MF& 0.3737 & 0.4450 & 0.1143 & 0.1275 & 0.2074 & 0.2622 & 0.1933 &0.1054 
\\
 MF-Random& 0.4122(+0.0385)& 0.4714(+0.0264)& 0.1434(+0.0291)& 0.1484(+0.0209)& 0.2618(+0.0544)& 0.3422(+0.0800)& 0.2302(+0.0369)&0.1279(+0.0225)
\\
 MF-Assistant& 0.4300(+0.0563)& 0.4781(+0.0331)& 0.1520(+0.0377)& 0.1593(+0.0318)& 0.2998(+0.0924)& 0.3706(+0.1084)& 0.2651(+0.0718)&0.1487(+0.0433)\\
\midrule
 ENMF& 0.4320 & 0.3953 & 0.0994 & 0.0997 & 0.0652 & 0.1036 & 0.2630 &0.1227 
\\
 ENMF-Random& 0.4931(+0.0611)& 0.4544(+0.0591)& 0.1195(+0.0201)& 0.1199(+0.0202)& 0.0751(+0.0099)& 0.1156(+0.0120)& 0.3056(+0.0426)&0.1446(+0.0219)
\\
 ENMF-Assistant& 0.5200(+0.0880)& 0.4831(+0.0878)& 0.1224(+0.0230)& 0.1217(+0.0220)& 0.0788(+0.0136)& 0.1247(+0.0211)& 0.3224(+0.0594)&0.1531(+0.0304)\\
\midrule
 NeuMF& 0.4720 & 0.4878 & 0.1364 & 0.1385 & 0.2160 & 0.2704 & 0.2891 &0.1507 
\\
 NeuMF-Random& 0.4464(-0.0256)& 0.4517(-0.0361)& 0.1559(+0.0195)& 0.1578(+0.0193)& 0.3301(+0.1141)& 0.3913(+0.1209)& 0.3220(+0.0329)&0.1603(+0.0096)
\\
 NeuMF-Assistant& 0.4856(+0.0136)& 0.4906(+0.0028)& 0.1631(+0.0267)& 0.1658(+0.0273)& 0.3507(+0.1347)& 0.4086(+0.1382)& 0.3451(+0.0560)&0.1742(+0.0235)\\
\midrule
 ItemKNN& 0.1211 & 0.1035 & 0.0889 & 0.0694 & 0.2242 & 0.3074 & 0.1657 &0.0790 
\\
 ItemKNN-Random& 0.1900(+0.0689)& 0.1698(+0.0663)& 0.1326(+0.0437)& 0.1051(+0.0357)& 0.2500(+0.0258)& 0.3035(-0.0039)& 0.2338(+0.0681)&0.1090(+0.0300)
\\
 ItemKNN-Assistant& 0.2131(+0.0920)& 0.1860(+0.0825)& 0.1517(+0.0628)& 0.1171(+0.0477)& 0.2660(+0.0418)& 0.3125(+0.0051)& 0.2567(+0.0910)&0.1170(+0.0380)\\
\bottomrule
\end{tabular}
\end{table*}

\subsection{Expansion Experiments of Burden Reduction}
In our Section~\ref{sec:reduce_human_burden}, we have compared the assistant's generation of feedback on behalf of users in the Proxy Set, and then passed this feedback to the recommendation system to help users further optimize the recommendation system. From our previous results, it can be seen that, with limited user interaction history and after learning about the user's personality, the assistant can effectively act on behalf of the user, optimizing various recommendation systems while reducing repetitive user operations. However, there might be a potential issue that predicting on the user's Proxy Set could leak the data distribution. Therefore, we conducted additional experiments to investigate whether the assistant truly helps in reducing the user's burden.

In Table~\ref{table:expand_e2}, we included an additional experiment: we used a program that randomly decides whether to like or dislike to simulate a non-intelligent assistant. Experimental results show that even randomly guessing likes and dislikes on the proxy dataset can improve the effect of the recommendation system in most experiments, indicating potential data distribution leakage risks. However, overall, the assistant designed based on our method outperformed the random program. This further validates our findings that the assistant can indeed be relatively intelligent to help users more easily optimize the recommendation system through proxy feedback.

\appendix

\end{document}